\documentclass[epj]{svjour}

\usepackage{graphicx}
\usepackage{fancyhdr}
\def\makeheadbox{{
 }}
\def\mytitle{My title} 
\def\myauthors{My name}  
\def\mytype{My type of session}
\def\mysession{My session}

\def\mytitle{Sensitivity of the LHC Experiments to Extra Dimensions} 
\def\myauthors{Dr Tracey Berry}    
\def\mytype{Contributed Talk}    
\def\mysession{Alternatives}

\begin{document}
\title{Sensitivity of the LHC Experiments to Extra Dimensions}
\author{Dr Tracey Berry\inst{1}
}                     
%
%
\institute{Royal Holloway, University of London, UK
}
%
\date{}
\abstract{
This conference report briefly reviews the potential of the ATLAS and CMS experiments to discover evidence of extra dimensions. 
%
\PACS{
      {PACS-key}{Extra Dimensions}   \and
      {PACS-key}{LHC}
     } 
} 
\maketitle
%

\section{Introduction}
\label{intro}
The Standard Model (SM) been very successful in describing physical phenomena up to energies $\sim$~100~GeV. Despite this, some questions remain unanswered: for example: Why are there three types of quarks and leptons of each charge? Is there some pattern to their masses? Are the quarks and leptons really fundamental, or do they have substructure? What particles form the dark matter in the universe? How can the gravitational interactions be included in the SM? In order to solve these issues, many theories to go beyond the SM have been proposed. In particular attempts to address the hierarchy problem (why the electroweak scale (1 TeV) is much smaller than the Planck scale (10$^{19}$GeV)) have resulted in the introduction of extra dimensional (ED) models. Interestingly, many of these predict results that can be experimentally detected at high energy detectors, such as ATLAS and CMS at the LHC. 
\par In this paper, the sensitivity of the LHC experiments to the ADD model~\cite{ADD}, RS~\cite{RS} and TeV$^{-1}$~\cite{TeV} is discussed with consideration to the observable signatures and the discovery potential of the ATLAS and CMS experiments.

\section{ADD Model}
\label{sec:1}

\par The ADD model has many large compactified extra dimensions, in which gravity can propagate~\cite{ADD}. In contrast, the SM particles are confined to a brane. In this model, the effective ($M_{Pl(4+n)}$) and the original Planck scale ($M_{Pl}$) are related by the equation: $M_{Pl}^{2}$ $\sim$ $R^{n}M_{Pl(4+n)}^{(2+n)}$, where $n$ is the number of extra dimensions. In order to address the hierarchy problem, the effective Planck scale can be reduced to the order of a TeV, if the size of the extra dimensions is large. (Assuming toroidal shaped extra dimensions: $R$$\sim$10$^{(30/n-19)}$ m, so for $n$$\geq$2, R $<$ 1mm.)

In the ADD model there are two ways in which the model could be experimentally detected: either via direct graviton emission in association with a vector-boson (a photon or a jet) or via virtual graviton exchange. 
In the first case, existence of the emitted graviton is deducible by a signature of missing transverse energy (MET) in a detector. Virtual graviton exchange could potentially be detected by deviations in dilepton and dijet cross-sections from those expected from only SM processes. In the ADD model, a broad change in cross-section at large invariant mass is expected, due to the summation over the closely spaced Kaluza-Klein (KK) towers of the graviton.
The discovery potential for both of these cases has been investigated at the LHC experiments.



\subsection{Graviton Emission Searches: $\gamma$ + MET}
\label{sec:1sub1} 

\begin{figure}
\includegraphics[width=0.45\textwidth,height=0.25\textwidth,angle=0]{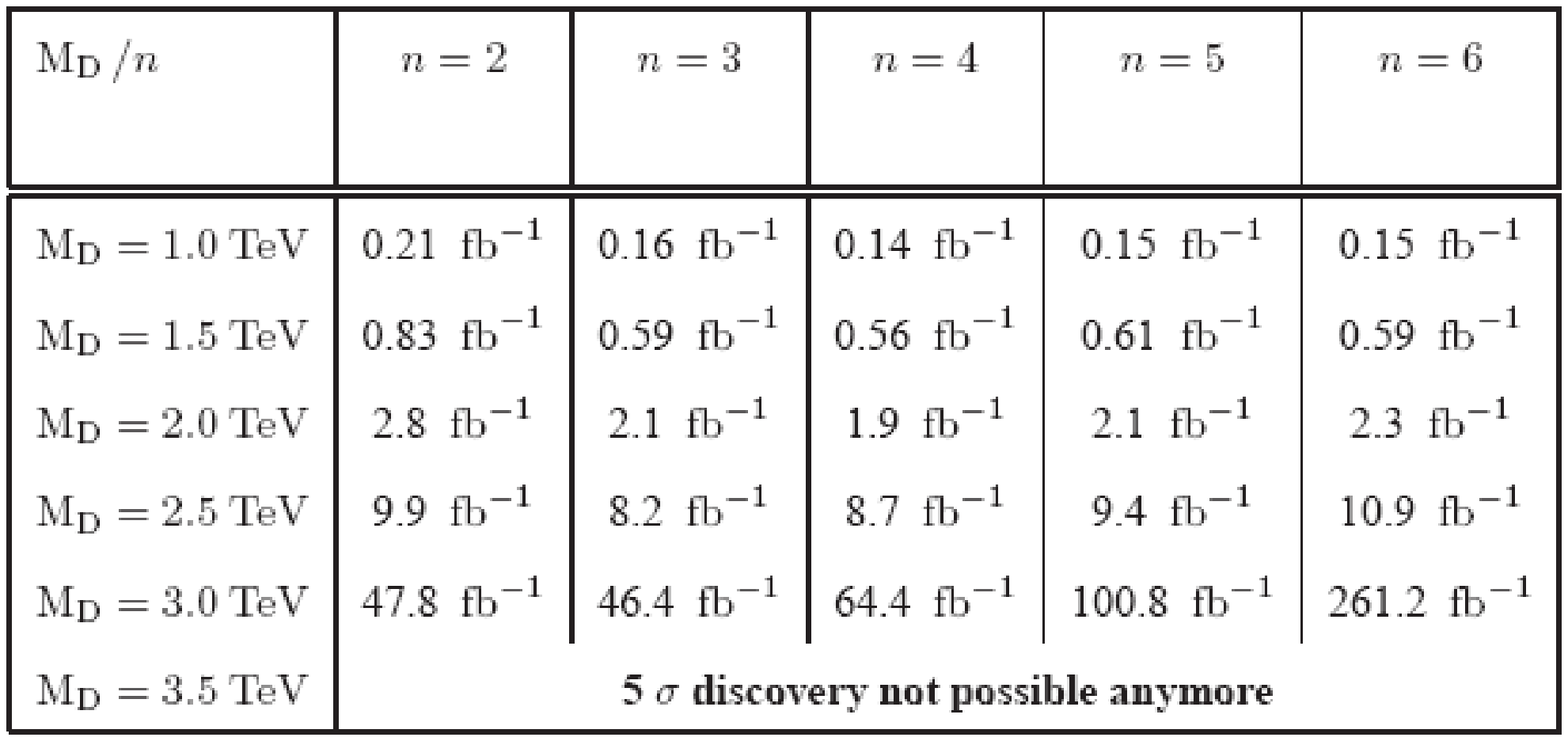}
\caption{Integrated luminosity required by the CMS experiment for a 5 $\sigma$ significance discovery in the $\gamma$+MET channel in searches for ADD model: shown as function of the gravitational scale (M$_D$) and the number of the extra dimensions (n)~\cite{ADDPhotonMetCMS}.}
\label{fig:ADDEmGamTable}       
\end{figure}

The discovery limit for searches has been investigated using the $\gamma$+MET channel by both the ATLAS and CMS collaborations. CMS searched for events which contained a photon with high-transverse momentum ($p_{T}$) and which had a large MET~\cite{ADDPhotonMetCMS}. The main background to the search is from the irreducible Z$\gamma$$\rightarrow$$\gamma$$\nu$$\nu$. (This is so large at low values of $p_{T}$ that the analysis requires that the photon has $p_{T}$$>$400 GeV/$c$). Other backgrounds included in the analysis are from $W$$\rightarrow$$e$($\mu$,$\tau$)$\nu$, $W$$\gamma$$\rightarrow$$e$$\nu$$\gamma$, $\gamma$+jets, QCD, di-photons and $Z^{0}$+jets. The estimated background rate from cosmic rays, which was the largest background in the CDF search~\cite{CDFADDphot}, has not been included in this analysis. The luminosity required for a 5$\sigma$ discovery using a significance definition of $S$ = 2($\sqrt{(S+B)}-\sqrt{B}$), and requiring $S$$>$5 is displayed in Figure~\ref{fig:ADDEmGamTable}. These are calculated using full simulation and reconstruction. Note that  5$\sigma$ discovery is not posssible for $M_{D}$ ($\sim$$M_{Pl(4+n)}$) $>$ 3.5 TeV. This is because the rates would be too low to observe. 


The ATLAS experiment is also sensitive to this discovery channel. For high luminosity running and with 100 fb$^{-1}$ of data the reach is up to 4 TeV for 2 extra dimensions~\cite{ADDPhotonMetATLAS}. More constraining limits come from searches in the jet+MET channel.


\subsection{Graviton Emission Searches: Jet + MET}
\label{sec:ADDEmission}

\begin{figure}
\begin{center}
\includegraphics[width=0.35\textwidth,height=0.3\textwidth,angle=0]{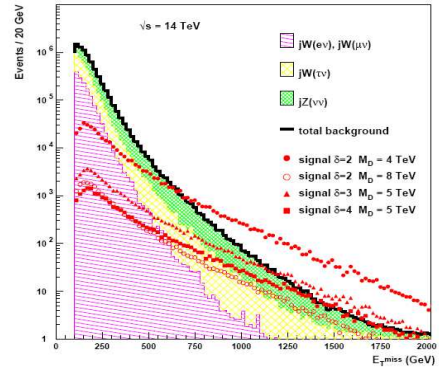}
\caption{Missing transverse energy distribution for jets+MET channel predicted at ATLAS in the ADD model~\cite{ADDPhotonMetATLAS}.}
\label{fig:2ADDGEmJetTT}       
\end{center}
\end{figure}

The ATLAS search for graviton emission requires a jet with high transverse momentum ($>$ 500 GeV) and a high missing transverse energy ($>$ 500 GeV) (from the escaping gravitons)~\cite{ADDPhotonMetATLAS}. The main backgrounds to the search are from the irreducible jet+$Z$$\rightarrow$jet+$\nu$$\nu$, which can be estimated using the $ee$ and $\mu\mu$ decays of the $Z$, and from jet+$W$$\rightarrow$jet+$l$$\nu$. In order to reduce the jet+$W$ background, leptons are vetoed. The contribution from these backgrounds and the total background are shown as a function of missing tranverse energy in Figure~\ref{fig:2ADDGEmJetTT}. Also shown is the signal for several values of the model parameters: the four dimensional Planck scale (M$_{D}$) and the number of extra dimensions ($\delta$).
The maximum reach in $M_{D}$ for low luminosity running and 30 fb$^{-1}$ is 7.7, 6.2, 5.2 TeV corresponding to 2, 3 and 4 extra dimensions. For high luminosity running and 100 fb$^{-1}$ these results can be extended up to 9.1, 7.0 and 6.0 TeV respectively. 

It is interesting to consider that if an excess in the number of events were observed, in order to characterise the model, both $M_{D}$ and the number of extra dimensions ($\delta$) would need to be determined. Measuring the shape of the MET spectrum can give ambiguous results. For example, the cross-section (at 14 TeV) in the case of $\delta$=2 and M$_{D}$=6~TeV is similar to that predicted for $\delta$=3 and M$_{D}$=5~TeV (Figure~\ref{fig:2ADDGEmJetTT}). A method to distinguish these two cases has been proposed which involves operating the LHC collider at two different centre of mass energies ($\sqrt{s}$) and studying the ratios of the predicted cross-section for runnning at 10 TeV and 14 TeV~\cite{ADDPhotonMetATLAS}.

%

%

\subsection{Graviton Exchange Searches: $ee$, $\mu\mu$, $\gamma\gamma$}
\label{sec:ADDExchange}

\begin{figure}
\begin{center}
\includegraphics[width=0.39\textwidth,height=0.3\textwidth,angle=0]{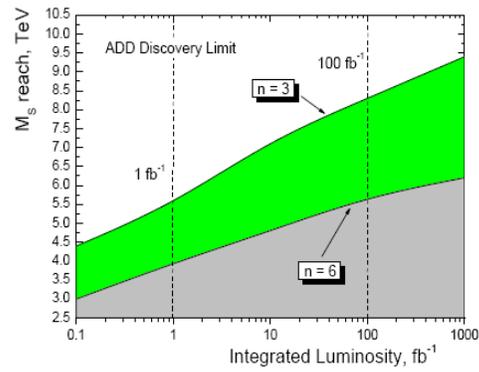}
\end{center}
\caption{CMS ADD graviton exchange discovery limits~\cite{CMSmmADDEx}.}
\label{fig:3ADDEx}       
\end{figure}

\begin{figure}
\includegraphics[width=0.45\textwidth,height=0.27\textwidth,angle=0]{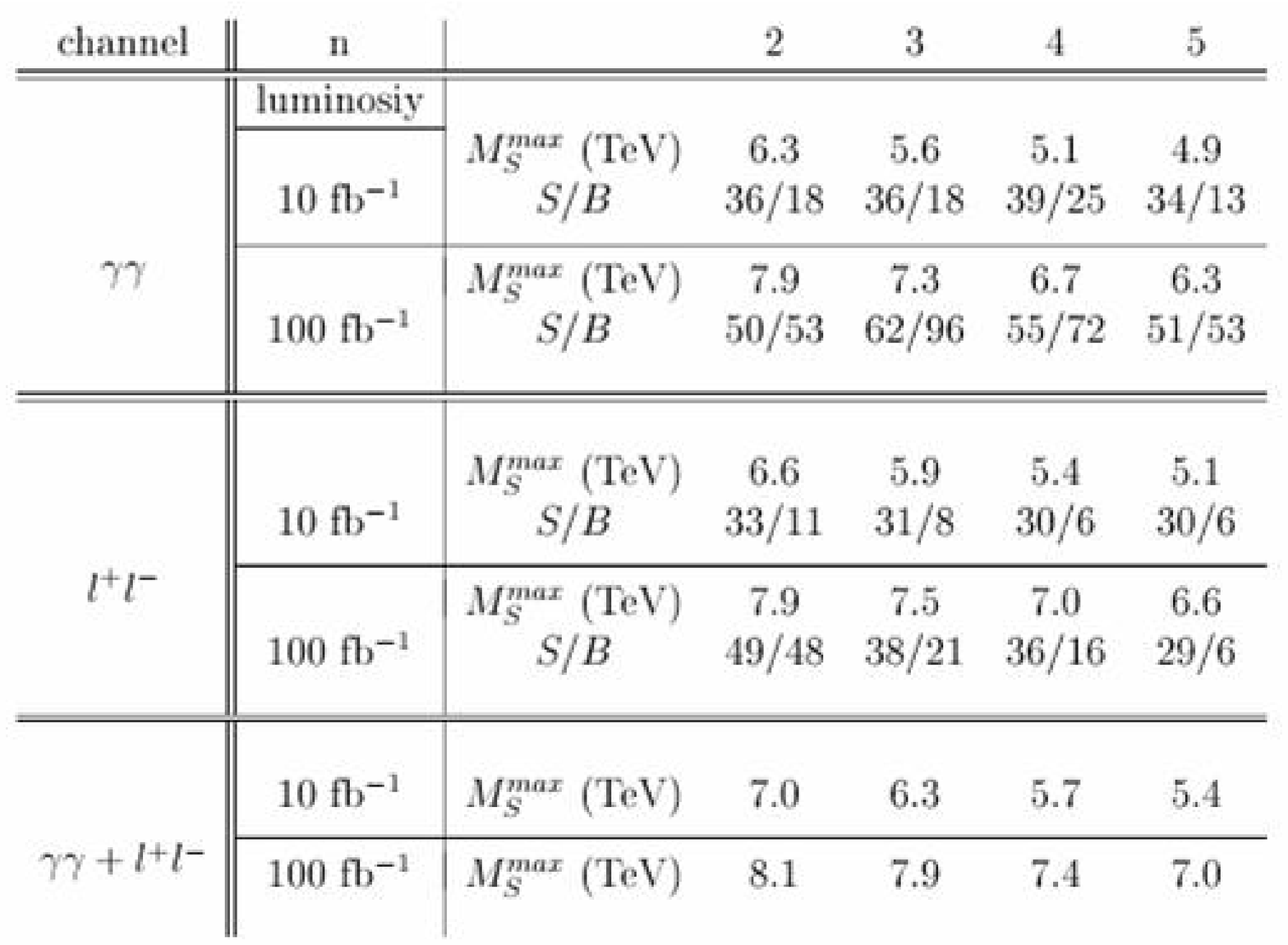}
\caption{Summary of predicted reaches achievable using ATLAS data, for the diphoton, dilepton and combined channel as a function of the number of extra dimensions (n) for the ADD model. Also shown are the predicted signal divided by background ratios (S/B) for each channel~\cite{ATLASADDExc}.}
\label{fig:ADDExchangeTable}       
\end{figure}

More exclusive limits on the ADD model come from searches for graviton exchange. CMS has studied the reach achievable in the dimuon channel~\cite{CMSmmADDEx}, requiring two opposite sign muons which have an invariant mass greater than 1 TeV. The background to this search is predominantly from the irreducible Drell-Yan contribution, but also from $ZZ$, $WW$, and $t\bar{t}$ events. The discovery reach achievable against the luminosity required is displayed in Figure~\ref{fig:3ADDEx} for 3 to 6 dimensions. The reach obtained with only 1 fb$^{-1}$ ranges from 3.9 to 5.5 TeV for 6 to 3 extra dimensions respectively. For 100 fb$^{-1}$ the range goes up to 5.7 to 8.3 and with 300  fb$^{-1}$ it is possible to probe up to 5.9 to 8.8 TeV.
Figure~\ref{fig:ADDExchangeTable} shows the reach achievable in ATLAS for an integrated luminosity of 10 fb$^{-1}$ and 100 fb$^{-1}$ for the diphoton and dilepton channels and also their combined results. Similarly to CMS, ATLAS can reach up to $\sim$5 TeV with 10 fb$^{-1}$ and up to $\sim$7 TeV with 100~fb$^{-1}$ in the dilepton channel~\cite{ATLASADDExc}. 




%

\section{RS Model}
\label{sec:RS}

 In the RS model, the hierarchy is solved by having a single highly curved (warped) extra dimension. In this scenario the SM particles are located on a brane in the extra dimension. The scale of new phenomena ($\Lambda_\pi$), in this model, is related to the Planck scale via the equation: $\Lambda_\pi$= $M_{Pl}e^{-kR_{c}\pi}$. Therefore, $\Lambda_\pi$ can be reduced to $\sim$1~TeV, if the curvature of this extra dimension is such that $kR_{c}$ $\sim$ 11-12. One of the experimental signatures for this model is a narrow high mass resonance in the dilepton, diboson or dijet channel. 

\subsection{Graviton Resonance Searches: $ee$, $\mu\mu$, $\gamma\gamma$}
\label{sec:RSGravExch}
\begin{figure}
\begin{center}
\includegraphics[width=0.39\textwidth,height=0.27\textwidth,angle=0]{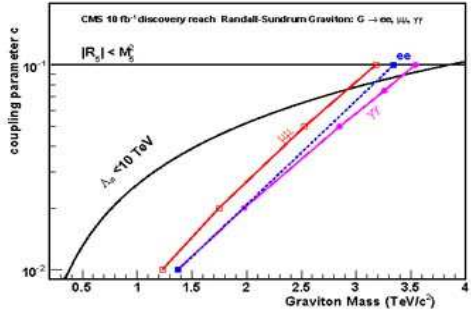}
\end{center}
\caption{Exclusion plane for the RS model achievable using 10 fb$^{-1}$ of CMS data. The region with c$>$0.1 is disfavoured as the bulk curvature becomes to large (larger than the five-dimensional Planck scale)~\cite{CMSRS}.}
\label{fig:TRSAllChan}       
\end{figure}

\begin{figure}
\begin{center}
\includegraphics[width=0.39\textwidth,height=0.27\textwidth,angle=0]{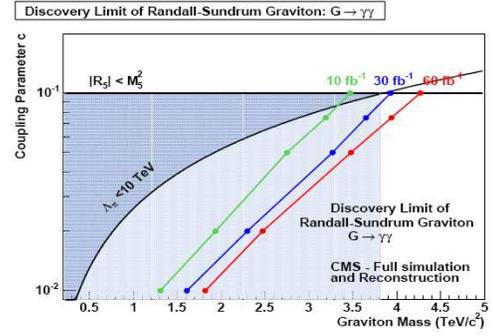}
\end{center}
\caption{Exclusion region of the RS model achievable using diphoton data at CMS. The region of interest is completely covered with 60 fb$^{-1}$ of data~\cite{CMSRS}.}
\label{fig:HRSgg}       
\end{figure}

The RS model has been studied by both ATLAS and CMS. At ATLAS the best channels to search in are the dielectron and diphoton channel due to the energy and angular resolutions of its detectors~\cite{Allenach02}. This is also confirmed by the CMS results using 10 $fb^{-1}$ shown in Figure~\ref{fig:TRSAllChan}~\cite{CMSRS}. This figure shows the exclusion plane of coupling parameter (c) against the mass of the graviton ($M_{G}$) (first resonance state). At low values of the coupling parameter and $M_{G}$, the results for the electron and photon are comparable, despite the branching ratio of the graviton to diphotons being twice that of the graviton to dimuons/dielectrons. This is because the background from QCD and prompt photons in the diphoton channel is harder to suppress efficiently, so for low values of $M_{G}$ the dielectron channel is as exclusive as the diphoton channel. The dimuon channel trails the dielectron channel due to the detector resolutions. At higher values of the $M_{G}$ the diphoton channel becomes more exclusive. Figure~\ref{fig:HRSgg} shows the reach for 10, 30 and 60~fb$^{-1}$ in the diphoton channel at CMS~\cite{CMSRS}. It is interesting that with less than 60~fb$^{-1}$ of data the region of interest (dark shaded area in Figure~\ref{fig:HRSgg}) can be completely covered. 


%

Should a resonance be observed, ATLAS could determine a spin-2 from a spin-1 resonance at the 90 \% confidence level for a graviton mass of up to 1.7 TeV with 100 fb$^{-1}$~\cite{Allenach02}.

\section{TeV$^{-1}$ Model}
\label{sec:TeV}
The final model covered in this overview is the TeV$^{-1}$ size extra dimensional model. 

In the models considered only fermions are confined to a brane within the extra dimensions, whereas the gauge fields ($W$, $Z$, $\gamma$ and g) can propagate into the extra dimensions. Consequently, this results in KK excitations of the gauge bosons of which the first could potentially produce observable resonances in a detector. In addition, detection of this model may be possible via the mixing of the zeroth mode (SM gauge boson) and the nth modes (n=1,2,3..) of the $W$ and $Z$ bosons and consequent interference phenomena. 



The TeV$^{-1}$ model could lead to detectable signatures in the invariant mass spectra of dileptons with deviations produced from the KK modes of the $\gamma$ and $Z^{0}$~\cite{ATLASTeV}~\cite{CMSTeVee}; also in the lepton-neutrino transverse invariant mass spectrum from the KK modes of the $W^{+/-}$~\cite{ATLASTeVW}; or alternatively via evidence of gluon resonances causing deviations in the dijet cross-section, or $b$$\bar{b}$ or $t$$\bar{t}$ cross-sections (not covered here)~\cite{TeVbbtt}. 

\subsection{$Z_{KK}^{1}$/$\gamma_{KK}^{1}$}
\label{sec:ZKK}

The discovery potential for the first KK resonance of the Z or $\gamma$ ($Z^{(1)}_{KK}$/$\gamma^{(1)}_{KK}$) has been investigated by ATLAS in both the dimuon and the dielectron channel. The resolving power for electrons at ATLAS is predicted to be better than that for muons, therefore more exclusive results are found in the dielectron channel and this channel would be used for discovery. 
It has been predicted that at ATLAS it would be possible to detect a resonance in the lepton-lepton ($ee$+$\mu\mu$) invariant mass spectrum of up to an mass of 5.8 TeV with 100 fb$^{-1}$~\cite{ATLASTeV}. This is comparable to the CMS result of 5.5/6 TeV for 30/80~fb$^{-1}$ for CMS to detect a peak in the dielectron spectrum~\cite{CMSTeVee}.
By searching for interference in a mass window (1000-2000 GeV) (rather than looking for a resonance) it is possible to increase the discovery reach achievable to $\sim$8 TeV for an integrated luminosity of 100 fb$^{-1}$ and $\sim$10.5~TeV for 300 fb$^{-1}$. In an optimised model specific search, if no peak is observed, the limit on the compactification scale ($M_{C}$) could be extended up to $\sim$ 13.5 TeV (300 fb$^{-1}$).



Studies have also been made on methods to distinguish TeV$^{-1}$ size ED model resonances from other models which predict resonances by studying the angular distributions and the forward-backward asymmetries~\cite{ATLASTeV}.

\begin{figure}
\begin{center}
\includegraphics[width=0.3\textwidth,height=0.3\textwidth,angle=0]{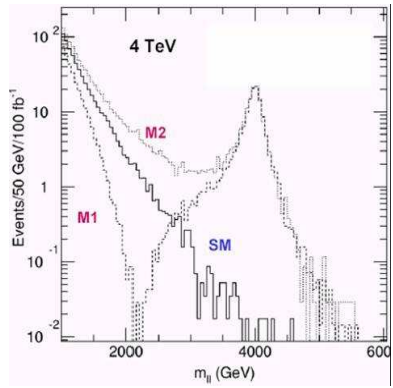}
\end{center}
\caption{Invariant mass distribution of $e^{+}e^{-}$ pairs for the SM (full line) and for model M1 (dashed line) (all of the SM fermions are on the same orbifold point) and M2 (dotted line) (quarks and leptons are located at opposite fixed points in the orbifold) in which the mass of the lowest lying KK excitation is 4000 GeV. The histograms are normalized to 100 fb$^{-1}$~\cite{ATLASTeV}.}
\label{fig:TeV}       
\end{figure}

\subsection{$W_{KK}$}
\label{sec:WKK}
ATLAS is also sensitive to searches for TeV$^{-1}$ ED via the KK modes of the $W$ boson, by searching in the lepton-neutrino transverse invariant mass spectra ($m_{T}^{l\nu}$)~\cite{ATLASTeVW}. By searching for a peak with 100 fb$^{-1}$ of data it is possible to set a limit on the compactification scale of $<$ 6~TeV for a combination of the electron and muon channel. Alternatively, a higher limit of 11.7~TeV for 100 fb$^{-1}$ can be ascertained by searching for interference effects below the peak in the $m_{T}^{e\nu}$ spectra, rather than search for the resonance itself. 

\begin{figure}
\begin{center}
\includegraphics[width=0.31\textwidth,height=0.28\textwidth,angle=0]{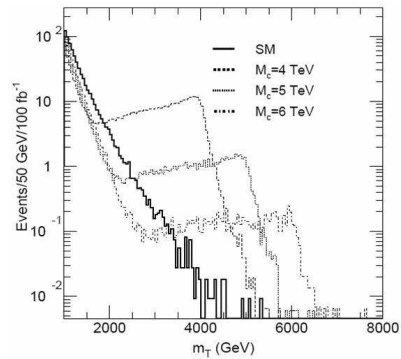}
\end{center}
\caption{Transverse invariant mass spectrum of $e\nu$ showing SM (solid line) spectrum and the distributions predicted for a compactification scale of 4, 5 and 6 TeV in the TeV$^{-1}$ model~\cite{ATLASTeVW}.}
\label{fig:wtenu}       
\end{figure} 
%




\section{Conclusion}
\label{sec:Conclusion}
The discovery potential of both experiments makes it possible to investigate if extra dimensions really exist within various ED scenarios at a few TeV scale. 
The reaches in the different channels will depend on the performance of the detector subsystems.
It is an exciting prospect that with a relatively small integrated luminosity ($\sim$ 10 fb$^{-1}$) there is potential to discover evidence of extra dimensions or to exceed the existing limits. For example: with 10 fb$^{-1}$ of data the reach on the fundamental Planck scale in the ADD model can be increased to $\sim$7-5 TeV for 2 to 6 extra dimensions (using the $\gamma\gamma$ and $l^{+}l^{-}$ channel searches); the discovery limit on the graviton mass in the RS model can be increased up to $\sim$3.5 TeV ($\gamma\gamma$ channel) and the compactification scale in the TeV$^{-1}$ model can probed up to $\sim$5.5 TeV ($e^{+}e^{-}$ channel).


\section{Acknowledgements}
\label{sec:Acknowledgments}
The author would like to thank the organisers of the SUSY 2007 conference for inviting her to talk and also to thank the ATLAS and CMS collaborations.

%

\begin{thebibliography}{999}
%
%


\bibitem{ADD} N. Arkani-Hamed, S. Dimopoulos and G. Dvali, Phys. Lett. \textbf{B429} (1998) 263; Nuc. Phys. \textbf{B544} (1999).  

\bibitem{RS} L. Randall and R. Sundrum, Phys. Rev. Lett. \textbf{83} (1999) 3370; Phys. Rev. Lett. \textbf{83} (1999) 4690.

\bibitem{TeV} Dienes, Dudas, Gherghetta, Nucl. Phys. \textbf{B537} (99); I. Antoniadis, PLB\textbf{246} (1990) 377.

\bibitem{ADDPhotonMetCMS} J.Weng {\it et al.}, CMS NOTE \textbf{2006/129}.

\bibitem{CDFADDphot} CDF Collaboration (D. Acosta {\it et al.}), Phys. Rev. Lett. \textbf{89} (2002) 281801.

\bibitem{ADDPhotonMetATLAS} L.Vacavant, I.Hinchcliffe, SN-ATLAS \textbf{2001-005}: J. Phys., \textbf{G 27} (2001) 1839-50.


\bibitem{CMSmmADDEx} I. Belotelov {\it et al.}, CMS NOTE \textbf{2006/076}; M. Lemaire {\it et al.}, CMS NOTE \textbf{2006/051}; CMS PTDR 2006.

\bibitem{ATLASADDExc} V. Kabachenko {\it et al.} ATL-PHYS-\textbf{2001-012}.

\bibitem{Allenach02} B. Allenach {\it et al.}, hep-ph 0211205; hep-ph 0006114.

\bibitem{CMSRS} I. Belotelov {\it et al.}, CMS NOTE \textbf{2006/104}; CMS PTDR 2006.




\bibitem{ATLASTeV} G. Azuelos, G. Polesello, SN-ATLAS \textbf{2003-023}; Eur. Phys. J. \textbf{C 39} (2005) 1-11.


\bibitem{CMSTeVee} B. Clerbaux {\it et al.}, CMS NOTE \textbf{2006/083}; CMS PTDR 2006b.

\bibitem{ATLASTeVW} G. Polesello, M. Patra, SN-ATLAS \textbf{2003-036}; G. Polesello, M. Patra, Eur. Phys. J.\textbf{C 32} (2004) 55-67.

\bibitem{TeVbbtt} L. March, E. Ros, B. Salvachua, ATL-PHYS-PUB-\textbf{2006-002}.


\end{thebibliography}
%

\end{document}